\documentclass[12pt]{article}
\begin{document}

\renewcommand{\thefootnote}{\fnsymbol{footnote}}

\begin{center}
{\bf \large RG/Pad\'e Estimate of the Three-Loop Contribution to the QCD
Static Potential Function}\\
\bigskip
F.A. Chishtie$^*$ and V. Elias\footnote{Permanent Address: Department of Applied Mathematics,
The University of Western Ontario,London, Ontario  N6A 5B7  CANADA}
\smallskip

Theory Group, KEK, Tsukuba, Ibaraki 305-0801, Japan
\end{center}

\bigskip

\renewcommand{\thefootnote}{\arabic{footnote}}
\setcounter{footnote}{0}

\begin{abstract}
The three renormalization-group-accessible three-loop coefficients 
of powers of logarithms within the $\overline{MS}$ series 
momentum-space for the QCD static potential are calculated and
compared to values obtained via asymptotic Pad\'e-approximant methods.
The leading and next-to-leading logarithmic coefficients are both found
to be in {\it exact} agreement with their asymptotic Pad\'e-predictions.
The predicted value for the third RG-accessible coefficient is found to
be within 7\% relative $|$error$|$ of its true value for $n_f \leq 6$, and
is shown to be in exact agreement with its true value in the $n_f
\rightarrow \infty$ limit.  Asymptotic Pad\'e estimates are also obtained 
for the remaining (RG-inaccessible) three-loop coefficient.
Comparison is also made with recent estimates of the three-loop contribution
to the configuration-space static-potential function.
\end{abstract}
The perturbative portion of the QCD static potential is presently known to two 
subleading orders of perturbation theory \cite{1,2,3}. This potential may be expressed as 
an integral over an $\overline{MS}$ perturbative-QCD series in momentum space,
\begin{equation}
V_{pert} (r)=\int\frac{d^3 q}{(2\pi)^3} e^{i\vec{q}\cdot \vec{r}} \left(
-\frac{16\pi^2}{3\vec{q}^{\;2}} \right) W \left[x(\mu),L(\mu,\vec{q}^{\;2})
\right]
\end{equation}
where
\begin{equation}
x(\mu)\equiv \alpha_s (\mu)/\pi, \; \; L(\mu, \vec{q}^{\;2})\equiv \log (\mu^2 / \vec{q}^{\;2}),
\end{equation}
and where the momentum-space series within the integrand of (1) is of the form
\begin{eqnarray}
W[x,L]=x[1+(a_0+a_1 L)x + (b_0+b_1 L+b_2 L^2)x^2 \nonumber \\
+ (c_0+c_1 L + c_2 L^2 + c_3 L^3)x^3 + ...]
\end{eqnarray}
with the following known coefficients \cite{1}:

\renewcommand{\theequation} {4\alph{equation}}
\setcounter{equation}{0}

\begin{equation}
a_0=31/12-5n_f / 18, \; \; a_1 = 11/4 - n_f / 6,
\end{equation}

\begin{equation}
b_0=28.5468-4.14714n_f + 25n_f^2 / 324,
\end{equation}

\begin{equation}
b_1=247/12 - 229n_f/72 + 5n_f^2 / 54,
\end{equation}

\begin{equation}
b_2=121/16-11n_f/12 + n_f^2 / 36,
\end{equation}
The three-loop order momentum-space coefficients $c_k$ have not been calculated.  
\footnote{A leading-log three-loop contribution in configuration space is extracted in refs. [4].}
  It should be noted, however, that the series' convergence may be problematical 
for values of $x$ near 0.1; {\it e.g.}, if $n_f=3$ and $\mu^2 = \vec{q}^{\;2}$,
$W[x,0] = x(1 + 1.75 x + 16.80 x^2  + ...]$.  There is clearly phenomenological 
value in having some knowledge of the next-order coefficients $c_k$ within the series (3), 
even if one chooses sufficiently large values of $\mu$ to ensure that the expansion 
parameter $x(\mu)$ remains small.

The all-orders momentum space static potential should ultimately 
be independent of the $\overline{MS}$ renormalization parameter $\mu$:

\renewcommand{\theequation} {\arabic{equation}}
\setcounter{equation}{4}

\begin{equation}
\mu^2\frac{d}{d\mu^2} W [x(\mu), L(\mu, \vec{q}^{\;2})] = 0.
\end{equation}
Eq. (5) corresponds to the following renormalization-group (RG) equation for the series $W[x,L]$:
\begin{equation}
\left( \frac{\partial}{\partial L} + \beta (x) \frac{\partial}{\partial
x} \right) W[x,L] = 0
\end{equation}
with \cite{5}
\begin{equation}
\beta(x) \equiv \mu^2 \frac{d}{d \mu^2} x(\mu) = - \sum_{k=0}^\infty
\beta_k x^{k+2},
\end{equation}

\renewcommand{\theequation} {8\alph{equation}}
\setcounter{equation}{0}

\begin{equation}
\beta_0 = 11/4 - n_f / 6,
\end{equation}
\begin{equation}
\beta_1 = 51/8 - 19 n_f / 24,
\end{equation}

\begin{equation}
\beta_2 = 2857/128 - 5033 n_f / 1152 + 325n_f^2 / 3456.
\end{equation}

To leading and next-to-leading orders in perturbation theory, Eq. (6) is manifestly 
satisfied by the known coefficients (4) for the perturbative series (3):

\renewcommand{\theequation} {\arabic{equation}}
\setcounter{equation}{8}

\begin{eqnarray}
0  & = & \left( \frac{\partial}{\partial L} + \beta \frac{\partial}{\partial x} \right) W[x,L] = (a_1 - \beta_0)x^2 \nonumber \\
& + & (b_1 - 2a_0 \beta_0 - \beta_1) x^3 + (2b_2 - 2\beta_0 a_1) x^3 L \nonumber \\
& + & {\cal{O}}(x^4, x^4 L, x^4 L^2).
\end{eqnarray}
The coefficients of $x^2, x^3$ and $x^3 L$ in (9) are all seen to vanish for known 
series coefficients (4) and $\beta$-function coefficients (8):  {\it i.e.}, the known values
of $b_1$ and $b_2$ are seen to uphold the RG-equation (6) by satisfying its perturbative
formulation (9).  However, it is 
important to note that (9) may also be utilized to extract all but one of the three-loop 
coefficients $c_k$ in the series (3).  The coefficients of $x^4, x^4 L$ and $x^4 L^2$ in 
(9) respectively vanish provided
\begin{equation}
c_1 = 290.769 - 60.4881 n_f + 3.2440 n_f^2 - 25 n_f^3 / 648,
\end{equation}
\begin{equation}
c_2 = 1639/16 - 4129 n_f / 192 + 377 n_f^2 / 288 - 5 n_f^3 / 216,
\end{equation}
\begin{equation}
c_3 = (11/4)^3 - 121n_f / 32 + 11 n_f^2 / 48 - n_f^3 / 216.
\end{equation}
Consequently, the only RG-inaccessible three-loop-order term in the series (3) is $c_0$. 
This coefficient can be obtained only via a direct perturbative
calculation, which has not yet been performed.

	In the absence of such a three-loop calculation, we employ the anticipated 
error of Pad\'e approximants in predicting next-order terms of a field theoretical series 
in order to obtain an estimate of all four three-loop coefficients
$c_k$ within (3).  The predictions for $\{c_1, c_2, c_3\}$ can then be compared to their 
true values (10-12) to check the validity of the estimation procedure.  The procedure 
we describe below has already been employed in a large number of applications: 
QCD $\beta$- and $\gamma$-functions \cite{6,7,8}, the SQCD  $\beta$-function \cite{7,9}, 
QCD current-correlation functions \cite{8,10}, the renormalization-group functions of 
O(N)-symmetric massive scalar field theory \cite{6,8,11}, Higgs decays \cite{8,10,12}, 
Higgs-mediated scattering processes \cite{13}, and QCD corrections to inclusive semileptonic B-decays \cite{14}. The 
general method we employ is described in refs. \cite{7}, \cite{8} and \cite{14}; we restate 
its development here for convenience.
  
Consider a perturbative series 
\begin{equation}
W(x)=1+R_1 x + R_2 x^2 + R_3 x^3 + ... + R_N x^N + ... \;  .
\end{equation}
For many such series, the series sum can be approximated by an $[N|M]$ Pad\'e-approximant, 
where N and M are respectively the degrees of numerator and denominator polynomials within 
the approximant. If only the next-to-leading term $R_1$ is known, for example, the Pad\'e
approximant
\begin{equation}
W^{[0|1]} (x) = \frac{1}{1-R_1 x} = 1 + R_1 x + R_1^2 x^2 + ...
\end{equation}
would predict a value of $R_1^2$ for the coefficient $R_2$ in (13).  Somewhat more 
realistically, if $R_1$ and $R_2$ in (13) are both known, the Pad\'e approximant
\begin{equation}
W^{[1|1]} (x) = \frac{1+(R_1-R_2/R_1)x}{1-(R_2/R_1)x} = 1+R_1 x + R_2
x^2 + (R_2^2/R_1) x^3 + ...
\end{equation}
leads to the predicted value $R_2^2/R_1$ for the coefficient $R_3$ in (13). Generally, 
one finds that the higher the degree of the approximant, the more accurate the prediction 
of the next unknown coefficient of the series will be. Suppose one now utilises an $[N-1|1]$ 
approximant to estimate the coefficient $R_{N+1}$ within (13), based upon knowledge of all 
previous series coefficients $\{R_1, R_2, .... , R_N\}$. For perturbative field-theoretical 
series, it is often found that the relative error in such an estimate is inversely 
proportional to N \cite{6,7,15}:
\begin{equation}
\left( R_{N+1}^{pred.} - R_{N+1}^{true} \right)/ R_{N+1}^{true} \cong -
A/N.
\end{equation}
The constant A in (16) can be estimated by comparing the $[0|1]$-approximant estimate for 
$R_2$ ({\it i.e.}, $R_2^{pred.} = R_1^2$) against $R_2$'s true value. One then finds 
via (16) that 
\begin{equation}
A \cong 1-R_1^2 / R_2.
\end{equation}
In the series (13), let us suppose we only know the subleading and NNLO 
coefficients $R_1$ and $R_2$, as is the case for the series (3).  If the $[1|1]$ approximant prediction $R_3^{pred.} = R_2^2/R_1$
has a relative error described by (16), we can substitute (17) into (16) to obtain the "true" 
value for $R_3$ {\it algebraically} \cite{8,14}:
\begin{equation}
R_3^{true} \cong \frac{R_2^2/R_1}{1-A/2} = \frac{2 R_2^3}{R_1^3 + R_1 R_2}.
\end{equation}

Of course, the validity of this result can be ascertained only by seeing how well it 
predicts coefficients that can be extracted by other means. \footnote{ The formula (18), for example, 
is surprisingly accurate in predicting the known four-loop order $\beta$-function coefficient 
in O(N)-symmetric massive scalar field theory \cite{13}.} 
For the case of the series (3), we identify the known coefficients $R_1$ and $R_2$ as 
polynomials in the logarithm L \cite{1}:
\begin{equation}
R_1 = a_0 + a_1 L = a_0 + \beta_0 L,
\end{equation}
\begin{equation}
R_2 = b_0 + b_1 L + b_2 L^2 = b_0 + (2a_0 \beta_0 + \beta_1) L +
\beta_0^2 L^2.
\end{equation}
Substituting (19) and (20) into (18), we obtain the following ``large-L'' series expansion for
$R_3$:  \footnote{Estimation of higher-order terms via such a series
expansion is denoted in ref. [10] as the ``APAP$^\prime$'' procedure.}
\begin{eqnarray}
R_3 & = & \beta_0^3 L^3 + (3a_0 \beta_0^2 + 5\beta_0 \beta_1
/ 2) L^2 \nonumber \\
& + & (a_0^2 \beta_0 / 2 + 5 \beta_0 b_0 / 2 + 5a_0 \beta_1 /
2 + 7 \beta_1^2 / 4 \beta_0 ) L^1 \nonumber \\
& + & \left[ \beta_0^3 (2D_1 D_2 - D_1^3 - D_3) + 3\beta_0 (2a_0
\beta_0 + \beta_1) (D_1^2-D_2) \right. \nonumber \\
& - & 3(2a_0 \beta_0 + \beta_1)^2 D_1/\beta_0 - 3\beta_0 b_0 D_1
\nonumber \\
& + & \left. (2a_0 \beta_0 + \beta_1)^3 / \beta_0^3 + 6b_0 (2a_0
\beta_0 + \beta_1)/\beta_0 \right] L^0 \nonumber \\
& + & {\cal{O}} (L^{-1}),
\end{eqnarray}
where

\renewcommand{\theequation} {22\alph{equation}}
\setcounter{equation}{0}

\begin{equation}
D_1 \equiv (6a_0 \beta_0 + \beta_1) / 2 \beta_0^2,
\end{equation}
\begin{equation}
D_2 \equiv [5a_0^2 \beta_0 + b_0 \beta_0 + a_0 \beta_1] /
2\beta_0^3,
\end{equation}
\begin{equation}
D_3 \equiv a_0 (a_0^2 + b_0)/2\beta_0^3.
\end{equation}
As is evident from (3), $R_3$ should be a degree-3 polynomial in the 
logarithm $L$.  A direct comparison of equivalent powers of $L$ in (3) 
and in (21) leads to the following predictions:
\renewcommand{\theequation} {\arabic{equation}}
\setcounter{equation}{22}
\begin{equation}
c_3^{pred.} = \beta_0^3,
\end{equation}
\begin{equation}
c_2^{pred.} = 3a_0 \beta_0^2 + 5\beta_0 \beta_1 / 2,
\end{equation}
\begin{equation}
c_1^{pred.} = a_0^2 \beta_0 / 2 + 5\beta_0 b_0 / 2 + 5\beta_1
a_0 / 2 + 7\beta_1^2 / 4 \beta_0,
\end{equation}
in addition to the predicted equivalence of the {\it unknown} coefficient $c_0$ with the 
lengthy square-bracketed term in (21).

The prediction (23) is in exact agreement with (12), the RG-determination of $c_3$, as is 
evident by substituting (8a) into (23).  Surprisingly, the predicted value (24) for $c_2$
is also in {\it exact} agreement with the RG value (11), as is evident from direct substitution 
of (8a), (8b) and (4a) into (24).  Note that this agreement for both coefficients is true 
for all values of $n_f$, indicating that the asymptotic error formula (16) replicates the 
RG-invariance of the series (3) to leading and next-to-leading order in the logarithm $L$, 
a most surprising result.

The formula (16) cannot, of course, replicate RG invariance to all orders in $L$, since 
the infinite series (21) which follows from it is not a degree-3 polynomial in $L$. Nevertheless, 
the coefficient of $L$ in (21) is strikingly close to the corresponding coefficient $c_1$ 
within (3), as obtained via RG-methods in (10).  In Table 1, such RG determinations of
$c_1$ are compared to the prediction (25).  As is evident from the Table, the predicted 
values for $c_1$ underestimate corresponding RG values by less than 7\% for $n_f \leq 6$, 
with the best agreement seen curiously to occur at $n_f = 6$. This feature may be understood 
by noting that the large-$n_f$ behaviour of the estimate (25),
\begin{equation}
c_1 
\begin{array}{c}{} \\
\longrightarrow \\
_{n_f \rightarrow \infty} \end{array}
- 25 n_f^3 / 648,
\end{equation}
is in exact agreement with that of (12), the RG-determination of $c_1$.

Table 1 also presents estimates of the coefficient $c_0$, as obtained from the 
(square-bracketed) $L^0$ term in (21). This coefficient, as noted earlier, cannot be 
extracted from lower-order terms via RG-methods.  It is nevertheless encouraging to note 
that corresponding predictions for $c_3$ and $c_2$ are exact, and that predictions of $c_1$
are nearly so. Thus, we have obtained in Table 1 asymptotic Pad\'e-approximant estimates for the three-loop
coefficient $c_0$, which, in conjunction with explicit $RG$-determinations (10-12) of the other three-loop
coefficients $\{c_1, c_2, c_3\}$ occuring within the perturbative series $W[x,L]$ (3),
constitute a prediction for the full three-loop contribution to the
static-potential integrand (1).

The coordinate-space potential corresponding to the series (3) can be obtained via (1) through use of the
following identities \cite{3}:

\renewcommand{\theequation} {27\alph{equation}}
\setcounter{equation}{0}

\begin{equation}
\int \frac{d^3 q}{(2\pi)^3} e^{i \vec{q} \cdot \vec{r}} \frac{1}{\vec{q}^{\; 2}} = \frac{1}{4\pi r},
\end{equation}

\begin{equation}
\int \frac{d^3 q}{(2\pi)^3} e^{i \vec{q} \cdot \vec{r}} \frac{\log (\mu^2/ \vec{q}^{\; 2})}{\vec{q}^{\; 2}}
= \frac{2 \left( \log  (\mu r) + \gamma_E \right)} {4\pi r},
\end{equation}

\begin{equation}
\int \frac{d^3 q}{(2\pi)^3} e^{i \vec{q} \cdot \vec{r}} \frac{ \left[\log (\mu^2/\vec{q}^{\; 2})\right]^2}{\vec{q}^{\; 2}}
= \frac{4 \left( \log (\mu r) + \gamma_E \right)^2 + \pi^2/3}{4\pi r},
\end{equation}

\begin{eqnarray}
& &\int \frac{d^3 q}{(2\pi)^3} e^{i \vec{q} \cdot \vec{r}} \frac{ \left[\log (\mu^2/\vec{q}^{\; 2})\right]^3}{\vec{q}^{\; 2}}\nonumber\\
\nonumber\\
& = & \frac{8 \left( \log (\mu r) + \gamma_E \right)^3 + 2\pi ^2 \left( \log (\mu r) + \gamma_E \right) + 16 \zeta (3)}
{4 \pi r}
\end{eqnarray}
where $\gamma_E = 0.577216$ and $\zeta(3) = 1.202057$.  Following ref [16], we set $\mu = 1/r$ and find that

\renewcommand{\theequation} {\arabic{equation}}
\setcounter{equation}{27}

\begin{equation}
V_{pert} (r) = \frac{\alpha_s (1/r)}{r} \sum_{n=0}^\infty V_n \alpha_s^n (1/r),
\end{equation}

\renewcommand{\theequation} {29\alph{equation}}
\setcounter{equation}{0}

\begin{equation}
V_0 = -4/3,
\end{equation}

\begin{equation}
V_1 = -(4/3 \pi) [a_0 + 2\gamma_E a_1],
\end{equation}

\begin{equation}
V_2 = -(4/3 \pi^2) [b_0 + 2\gamma_E b_1 + (4\gamma_E^2 + \frac{\pi^2}{3})b_2 ],
\end{equation}

\begin{equation}
V_3 = -(4/3 \pi^3) [c_0 + 2\gamma_E c_1 + (4\gamma_E^2 + \frac{\pi^2}{3})c_2  + \left( 8\gamma_E^3 + 2\pi^2 \gamma_E + 16 \zeta(3) \right)c_3].
\end{equation}
For arbitrary $n_f$, values of $\{a_0, a_1, b_0, b_1, b_2 \}$ are given by (4), and values of 
$\{c_1, c_2, c_3 \}$ are given by (10,11,12).  In Table 2 we display values of the coefficients
$V_{1-3}$ obtained via (29) for $n_f = \{3, 4, 5\}$.  The estimate for $V_3$ is obtained
through use of (10, 11, 12) and the estimated values for $c_0$ in the final column of Table 1.  In 
Table 2 we also list values of $V_3$ estimated via renormalon-matching (RM) considerations \cite{16}, as well as 
corresponding large-$\beta_0$ estimates of $V_3$ \cite{17}.  Striking agreement of all three estimation procedures
is clearly evident in Table 2.  However, it must be noted that $V_3$ is not very sensitive to $c_0$ [the only RG-inaccessible
coefficient in (29d)] when $\mu=1/r$.  If one uses (29d) to extract $c_0$ from $V_3$, for example, one finds that the 
$V_3$ values -38.4, -37.34 (the RM value), and -34.06 (large $\beta_0$) tabulated in Table 2 for $n_f=3$ respectively correspond to $c_0$ values of 142 (our
Table 1 RG/Pad\'e estimate), 116, and 40.

An alternative approach to estimating $c_0$ follows from a least-squares fit of the asymptotic
Pad\'e-approximant prediction (18) to the three-loop momentum-space contribution's
explicit dependence on $L$,

\renewcommand{\theequation} {\arabic{equation}}
\setcounter{equation}{29}

\begin{equation}
R_3 = c_0 + c_1 L + c_2 L^2 + c_3 L^3,
\end{equation}
over the entire ultraviolet ($\mu^2 > \vec{q}^{\;2}$) region, a procedure which has been 
employed previously in a number of different applications \cite{12,13,14}.  If we define 
$w \equiv \vec{q}^{\;2} / \mu^2$ [i.e., $\log(w) = -L$], such a procedure entails optimization of
\begin{equation}
\chi^2 [c_0] = \int_0^1 dw \left[ \frac{2 R_2^3 (w)}{R_1^3 (w) + R_1 (w) R_2(w)} - c_0 
+ c_1 \log (w) - c_2 \log^2 (w) + c_3 \log^3 (w)\right]^2,
\end{equation}
with respect to $c_0$, where
\begin{equation}
R_1(w) = a_0 - a_1 \log (w), R_2 (w) = b_0 - b_1 \log (w) +
b_2 \log^2 (w),
\end{equation}
and where the set of known coefficients $\{a_0, a_1, b_0, b_1, b_2, c_1, c_2, c_3\}$ is 
given by (4) and (10-12).  Unlike previous applications in which large-{\it L} expansions  
of (18) are quite consistent with least-squares fits, \footnote{For example, in
semileptonic $b \rightarrow u$ decay, the $n_f = 4$ $c_0$ coefficient has a large-{\it L}-expansion
value of 166, in approximate agreement with the estimate $c_0^{(4)} = 188$ obtained
in \cite{14} via (28) with known values \cite{18} for $\{a_0, a_1, b_0, b_1, b_2\}$ and RG-determinations
\cite{14} of $\{c_1, c_2, c_3\}$ appropriate for $b \rightarrow u \ell^- \bar\nu_{\ell}$.} such a fit 
is seen to lead to values of $c_0$ that are $\sim$50\% larger than those of Table 1:
\begin{equation}
n_f = 3: \; \; \chi^2 [c_0] = 42679 - 405.8 c_0 + c_0^2 \; \rightarrow
c_0 \; = 203,
\end{equation}
\begin{equation}
n_f = 4: \; \; \chi^2 [c_0] = 22142 - 291.1 c_0 + c_0^2 \; \rightarrow  \; c_0
= 146,
\end{equation}
\begin{equation}
n_f = 5: \; \; \chi^2 [c_0] = 9501 - 189.9 c_0 + c_0^2 \; \rightarrow \; c_0
= 95,
\end{equation}
\begin{equation}
n_f = 6: \; \; \chi^2 [c_0] = 2901 - 104.7 c_0 + c_0^2 \; \rightarrow \; c_0
= 52,
\end{equation}

In assessing the accuracy of (33-36), it should be noted that such least-squares fitting 
could also be employed to fit simultaneously {\it all four} three-loop coefficients 
$\{c_0, c_1, c_2, c_3\}$, as has been done before in a number of applications \cite{12,13,14} 
in which the fitted values for $\{c_1, c_2, c_3\}$ closely approximated their known RG values. 
However, such a procedure completely fails for the series (3):  when $n_f = 3$, optimization of
(31) with respect to $c_{0-3}$ yields values $[c_0 = 258, \; c_1 = 54.7,
\; c_2 = 66.3, \; c_3 = 10.2]$ that differ substantially from true
values $[c_1 = 137.46, \; c_2 = 49.078, \; c_3 = 11.391]$ obtained from
eqs. (10,11,12).  
A similarly large estimate of $c_0$ for the $n_f = 3$ case is obtained directly 
via (18) in the {\it L} = 0 (small-log) limit, in which $R_2 = b_0$ (4b) and 
$R_1 = a_0$ (4a). Such an approach yields $c_0 = 273$, a value quite comparable 
to that obtained above ($c_0 = 258$) by simultaneous least-squares fitting 
of all four three loop coefficients $c_{0-3}$. If one {\it inputs} this small-log 
estimate for $c_0 \; (= 273)$ into (31) and then minimizes with respect to the 
RG-accessible coefficients $c_{1-3}$, the estimated values for these 
coefficients will be even worse than those characterising the  
full least squares fit ($c_1 \simeq 33, \; c_2 \simeq 74, \;  c_3 \simeq 9.6$) -- values 
which are {\it inconsistent} (particularly $c_1$) with the RG determinations of these same 
parameters. 

Such discrepancies suggest that the large-$L$ ({\it i.e.} large $\mu$ or short distance) $c_0$
estimates of Table 1, which reproduce {\it exact} RG values for $c_2$ and $c_3$ and
closely approximate RG values for $c_1$, be taken more seriously than either  the $c_0$ 
estimates (33-36) obtained via least-squares fitting over a broad range of $\mu$, or other
({\it e.g.} small-log) approaches to estimating $c_0$ within an asymptotic Pad\'e-approximant context.
It is evident that such alternative asymptotic Pad\'e-approximant estimation procedures are of 
little value if not tied to some way of successfully estimating $c_{1-3}$, the RG-accessible three loop 
coefficients. By this criterion, the large-{\it L} estimates of Table 1 have the most substantial 
credibility. 

	However, it is important to remain cognisant of the relative insensitivity 
(noted earlier) of the configuration-space static potential to the parameter $c_0$ at 
its benchmark $\mu = 1/r$ length scale. At this scale, {\it RG-accessible coefficients 
alone} dictate that the $n_f = 3$ three-loop contribution $V_3$ is given by
\begin{equation}
V_3 = -4 \left[ c_0 + 752.0 \right] / 3 \pi^3,
\end{equation}
a result which follows from substitution of (10-12) into (29d).  Since all $n_f = 3$ 
estimates of $c_0$, as  delineated in the previous two paragraphs, are small compared 
to 752.0, $V_3$ is surprisingly insensitive to this unknown parameter; the four disparate 
estimates 142 [large-{\it L}], 203 [eq. (33)], 258 [full least-squares treatment], and 273 [small-log] all correspond to $V_3$ values 
near -40 [-38.4, -41.1, -43.4, and -44.1, respectively].  Thus, even if one trusts Pad\'e methods 
only to give a factor-of-two-accuracy estimate for the magnitude of $c_0$,  
RG-accessible coefficients {\it alone} are sufficient to extract a surprisingly 
concise range for the three-loop contribution $V_3$.  In other words, the $\mu = 1/r$ estimate 
for $V_3$, the three-loop quantity ultimately of phenomenological interest to us, is subject to 
substantially less theoretical uncertainty than the parameter $c_0$.
	
	A final and necessary {\it caveat}, however, is the possibility that new diagrammatic 
topologies (and their corresponding group theoretical factors) known to enter the QCD 
static potential at three-loop order \cite{4} may further circumscribe the applicability of 
using lower-order terms to predict the three-loop contribution $c_0$, as in a Pad\'e-approximant 
approach. This situation is entirely analogous to the theoretically uncertain light-by-light scattering 
contributions known to enter the muon's anomalous magnetic moment at sufficiently high 
order, as well as the quartic Casimir terms first appearing in the QCD $\beta$-function series 
at four-loop order.  Pad\'e-approximant based techniques cannot be expected to predict 
terms characterised by new higher-order group-theoretical factors \cite{7}. 
Nevertheless, such contributions do not necessarily dominate the 
first order in which they appear, nor do they necessarily devalidate Pad\'e estimates 
for that order. For example, the asymptotic Pad\'e-approximant estimate of the ($N_c=3$) four-loop 
contribution to the QCD $\beta$-function \cite{7} 
$\beta_3 =$ {\it 23600} - {\it 6400}$n_f +$ {\it 350}$n_f^2 +$ 1.49931$n_f^3$ (estimated numbers are italicised) 
is in quite reasonable agreement with the exact result \cite{19} 
$\beta_3$ = 29243.0 - 6946.30$n_f$ + 405.089$n_f^2$ + 1.49931$n_f^3$, though in much closer agreement 
with the calculated result with quartic Casimir terms excised \cite{7}: 
$\beta_3$ = 24633 - 6375$n_f$ + 398.5$n_f^2$ + 1.49931$n_f^3$. Thus, based on the limited  
information available, the remarkable success of 
the asymptotic Pad\'e-approximate large-{\it L} expansion in predicting those three-loop momentum-space static potential 
terms that are also extractable by RG methods encourages some confidence in
the corresponding large-{\it L} prediction of the 
RG-inaccessible parameter $c_0$ (modulo the above-mentioned uncertainties characterising Pad\'e approaches), 
at least for purposes of predicting $V_3$, the three-loop contribution to the {\it configuration-space} static potential.

\newpage

\section*{Acknowledgements}

We are grateful to A. Hoang for suggesting the application of 
Pad\'e-approximant methods to the QCD static potential and to N. Brambilla and
 A. Pineda for helpful correspondence.  We are also grateful to
the High-Energy Theory Division at KEK for providing hospitality during
the performance of this research, and to the Natural Sciences and Engineering 
Research Council of Canada for providing financial support for our stay in Japan.

\newpage

\begin{table}
\centering
\begin{tabular}{||c|c|c|c|c||}
\hline \hline
$n_f$ & $c_1^{RG}$ & $c_1^{Pred}$ & $|(c_1^{Pred} - c_1^{RG})/c_1^{RG}|$
& $c_0^{Pred}$ \\ \hline \hline
0 & 290.77 & 272 & 6.3\% & 313\\ \hline
1 & 233.49 & 218 & 6.5\% & 250\\ \hline
2 & 182.46 & 170 & 6.8\% & 193\\ \hline
3 & 137.46 & 128 & 7.0\% & 142\\ \hline
4 & 98.251 & 91.4 & 6.9\% & 97.5\\ \hline
5 & 64.606 & 60.6 & 6.2\% & 60.1\\ \hline
6 & 36.291 & 35.3 & 2.8\% & 30.5\\ \hline \hline
\end{tabular}
\caption{Comparison of predicted and RG values for the three-loop coefficient $c_1$.  Also
displayed are predicted values for the RG-inaccessible coefficient $c_0$.  Note that these
$c_0$ estimates are the same sign and approximate magnitude as the RG values for $c_1$ listed
in the second column.}
\label{Table I}
\end{table} 

\begin{table}
\centering
\begin{tabular}{||c|c|c|c|c|c||}
\hline \hline
$n_f$ & $V_1$ & $V_2$ & $V_3$ & $V_3^{RM}$ & $V_3^{L\beta_0}$\\ \hline \hline
3 & -1.84512 & -7.28304 & -38.4 & -37.34 & -34.06\\ \hline
4 & -1.64557 & -5.94978 & -28.7 & -27.63 & -27.03\\ \hline
5 & -1.44602 & -4.70095 & -20.5 & -19.46 & -21.05\\ \hline \hline
\end{tabular}
\caption{Configuration-space coefficients (28) of the configuration-space static
potential.  The column labeled $V_3$ is obtained using RG-determinations of $c_1, c_2, c_3$ 
and the Table 1 estimate of $c_0$ within eq. (29d).  The column labeled $V_3^{RM}$ is 
obtained from eq. (22) of ref. \cite{16}.  The column labeled $V_3^{L\beta_0}$ lists 
large-$\beta_0$ estimates \cite{17} that are also tabulated in ref. \cite{16}.}
\label{Table 2}
\end{table}
\end{document}